\documentstyle[11pt]{article}

\newcommand{\be}{\begin{equation}}
\newcommand{\ee}{\end{equation}}
\newcommand{\ba}{\begin{array}}
\newcommand{\ea}{\end{array}}
\newcommand{\bc}{\begin{center}}
\newcommand{\ec}{\end{center}}
\newcommand{\bi}{\begin{itemize}}
\newcommand{\ei}{\end{itemize}}

\newcommand{\disregard}[1]{{}}

\def\bild#1\over#2{\mathrel{\mathop{\kern0pt #1}\limits_{#2}}}

\begin{document}

\centerline {\bf MAGNETIC MOMENT and PERTURBATION THEORY }

\centerline {\bf with SINGULAR MAGNETIC FIELDS}
\vskip 1cm
\centerline {{\bf   Alain Comtet\rm}$^{(+)}$\footnote{\it  and
LPTPE, Tour 12, Universit\'e Paris  6},}

\centerline{{\bf Stefan Mashkevich\rm }$^{(++)}$ and
{\bf St\a'ephane Ouvry
\rm }$^{(+)1}$ }
\vskip 1cm
\centerline{{(+)Division de Physique Th\'eorique \footnote{\it Unit\a'e de
Recherche  des
Universit\a'es Paris 11 et Paris 6 associ\a'ee au CNRS},  IPN,
  Orsay Fr-91406}}
\centerline{(++) Institute for Theoretical Physics, Kiev 252143, Ukraine  }
\vskip 1cm
{\bf Abstract :}
The spectrum of a charged particle coupled to Aharonov-Bohm/anyon
gauge fields displays a nonanalytic behavior in the coupling constant.
Within perturbation theory, this gives rise to certain
singularities which can be handled by adding a
repulsive contact term
to the Hamiltonian.
We discuss the case of smeared flux tubes with
an arbitrary
profile and show that the
contact term can be interpreted as the coupling of a magnetic moment
spinlike degree of
freedom
to the magnetic field inside the flux
tube. We also clarify the ansatz for
the redefinition of the wave function.

\vskip 1cm

IPNO/TH 94-91

September 1994

 e-mail : COMTET@IPNCLS.IN2P3.FR, MASH@PHYS.UNIT.NO, OUVRY@FRCPN11

\vfill\eject

The fact that the problem of $N$ noninteracting anyons \cite{LM},
for $N>2$, is exactly solvable only in the two limit
cases of bosons and fermions,
gives rise to the idea of applying  perturbation theory
in order to get at least some
information in the vicinity of these two limit cases.
However, perturbation
theory meets certain difficulties near Bose statistics, as originally noticed
in \cite{first}.
In order to overcome these difficulties, it was pointed out in \cite{Ouvry92}
\cite
{Ouvry94}, that certain
modifications of the singular $N$-anyon Hamiltonian are required.

In the regular gauge,
anyons may be viewed \cite{Wilczek82} as charged particles
with attached singular Aharonov-Bohm \cite{AB} flux tubes.
In this letter,  we discuss a generalization of the
perturbative algorithm discussed in \cite{Ouvry92} \cite
{Ouvry94} to the case of smeared flux tubes with
any profile.
This  will bring some light on the singular case itself.
In particular, the contact repulsive interaction $\delta^2(\vec r)$ added
to the singular Hamiltonian will be reinterpreted as a magnetic moment
coupling of the particle to the magnetic field inside the flux tube.
Characteristic features of the singular perturbative algorithm, as for
example cancellation of singular 2-body as well as  regular 3-body interactions
in the transformed Hamiltonian,
will be shown to  be easily generalized provided that such magnetic moment
couplings are properly taken into account.

Let us first remind to the reader what happens in the paradigm
 Aharonov-Bohm (A-B) problem, or equivalently, in the
relative $2$-anyon
problem  \cite{Ouvry92} \cite{Ouvry94}. This is convenient since a
complete checking of the perturbative results
at all stages is possible for this problem by comparison against the exact
ones. We work in
the regular gauge, in which the wave functions
are single-valued and the A-B statistical parameter $\alpha$ explicitly
appears in the Hamiltonian. We consider a particle of charge $e$
and mass $m$ moving in a plane and coupled to the
gauge potential of a singular flux tube $\phi$ located at the origin
\be\label{1} H={1\over 2m}\left(\vec p-e \vec A
            \right)^2 \ee
where $\vec A(\vec r)= {\alpha\over e}
{\vec k\times\vec r\over r^2}$ and $\vec k $ is the unit vector
perpendicular to the plane.
 The A-B statistical parameter is $\alpha={e\phi\over 2\pi}$.
The Hamiltonian
(\ref{1}), in polar coordinates, is
\begin{equation}\label{Hamilt}
H ={1\over 2m}(-\frac{\partial^2}{\partial r^2}
- \frac{1}{r}\frac{\partial}{\partial r}
- \frac{1}{r^2}\frac{\partial^2}{\partial\phi^2}
+ \frac{2i\alpha}{r^2}\frac{\partial}{\partial\phi}
+ \frac{\alpha^2}{r^2})+ {1\over 2}m\omega^2r^2
\end{equation}
where one has added a harmonic attraction  in order
to
discretize the  spectrum.
The complete set of exact eigenstates
for this
Hamiltonian, up to a normalization,
is given by
\begin{eqnarray}
E_{\ell n} & = & (2n + |\ell-\alpha| + 1)\omega, \label{E} \\
\psi_{\ell n}(r,\phi) & = &  r^{|\ell-\alpha|}
{}_1 F_1 \left(-n,|\ell-\alpha|+1,
-{m\omega}r^2\right)
\exp\left( -{m\omega\over 2}r^2\right)
\exp\left( i\ell\phi \right)
\label{Psi}
\end{eqnarray}

The ground state wave function is obtained by setting
$\ell=0$, $n=0$
\begin{equation}\label{00}
\psi_{00}(r,\phi)= r^{|\alpha|}\exp\left(
-\frac{m\omega}{2}r^2\right)
\ee
Its energy is
\be E=(|\alpha|+1)\omega\label{E0}\ee
It is, however, impossible to get
(\ref{E0}) in perturbation theory near Bose statistics, treating the
$\alpha$ dependent terms in (\ref{Hamilt}) as perturbations.
Indeed, in the s-wave sector,  non-zero peturbative corrections
turn out to be logarithmically divergent. For example,
the unperturbed
ground state wave function $\psi^{(0)}_{00}=\sqrt{{m\omega\over
\pi}}\exp\left(
-\frac{m\omega}{2}r^2\right)$ gives
\begin{equation}
\left\langle \psi^{(0)}_{00} \right|\frac{\alpha^2}{2mr^2}
\left| \psi^{(0)}_{00}\right\rangle
= \int\limits_0^{\infty}\frac{\omega\alpha^2}{r}  e^{-m\omega
r^2}dr
\label{corr}
\end{equation}
The reason of
this divergence may be traced back to the fact that the unperturbed
$\ell=0$ wave function does not vanish at the origin
while the perturbed one does and therefore one cannot
get the latter as a perturbative series starting from the former.

In order to get a meaningful perturbation expansion, a modification
of the Hamiltonian is required.
Adding to  $H$
a short range repulsive interaction \cite{Ouvry94}, one defines
\begin{equation}\label{Hdelta}
H' = H + \frac{\pi|\alpha|}{m}\delta^2({\vec r})
\end{equation}
The  contact term clearly does not affect the exact wave functions, since
they  vanish
at the origin, except in the Bose case $\alpha=0$, but then there is no
contact
interaction. Still, this new Hamiltonian makes it possible
to use perturbation theory, with the parameter ${\pi|\alpha|\over m}$ in
(\ref{Hdelta}) precisely chosen for this aim. Indeed, the first order
correction to the ground state energy from the
contact term is just
\begin{equation}
\left\langle \psi^{(0)}_{00} \right|
\frac{\pi|\alpha|}{m}\delta^2({\vec r})
\left| \psi^{(0)}_{00}\right\rangle
= \frac{\pi|\alpha|}{m} \left| \psi^{(0)}_{00}(0) \right|^2
= |\alpha|\omega,
\label{exact}
\end{equation}
and it turns out that, while  the higher order corrections
due to this term are divergent, they nevertheless exactly
cancel the divergent corrections coming from the ${\alpha^2 \over r^2}$ term.
More precisely, the singular perturbative problem is solved in the sense that,
if a short range regulator is introduced to give an unambiguous
meaning to the perturbative divergences,  they do
cancel in the
limit where the regulator vanishes \cite{e}.

It would however be more satisfactory to have a perturbative algorithm
where perturbative divergences do not exist from the very beginning
\cite{Ouvry92}.
If, willing to take into
account the small $r$ behavior of the ground state
wave function (\ref{00}), one redefines \cite{Ouvry92}
\begin{equation}
\psi(r,\phi) = r^{|\alpha|}\tilde{\psi}(r,\phi)
\label{ansatz}
\end{equation}
then the Hamiltonian $\tilde {H}$ acting on $\tilde{\psi}$ no longer
contains  the dangerous
${\alpha^2\over r^2}$ term
\begin{equation}
\tilde{H} = {1\over 2m}(-\frac{\partial^2}{\partial r^2}
-\frac{1}{r}\frac{\partial}{\partial r}
-\frac{1}{r^2}\frac{\partial^2}{\partial\phi^2}
+\frac{2i\alpha}{r^2}\frac{\partial}{\partial\phi}
-\frac{2|\alpha|}{r}\frac{\partial}{\partial r})+ {1\over 2}m\omega^2r^2
\label{Htilde}
\end{equation}
The last term in the brackets in (\ref{Htilde}), which appears in place of the
singular
one, does not lead to any perturbative singularities. The first-order
correction to the ground state energy
\begin{equation}
\left\langle \psi^{(0)}_{00} \right| - \frac{|\alpha|}{mr}
\frac{\partial}{\partial r} \left| \psi^{(0)}_{00} \right\rangle=
|\alpha|\omega,
\label{exactpri}
\end{equation}
does coincide with the exact answer, while the higher
order perturbative corrections are finite and cancel. The fact
that first order perturbation theory gives here
the exact answer is of course due to the fact that one has
``guessed'' the correct ansatz (\ref{ansatz}) by
looking at the exact solution (\ref{00}).

In perturbation theory
for the $N$-anyon problem \cite{Ouvry92}, the ansatz analogous to
(\ref{ansatz}),
\begin{equation}
\psi = \prod\limits_{j<k} r_{jk}^{|\alpha|} \tilde{\psi},
\label{ansatzN}
\end{equation}
eliminates in $\tilde{H}$ not only the singular 2-body terms,
 but also the
3-body terms, thus considerably
simplifying the treatment.
This complete cancellation
 can be understood if one remarks that the prefactor
$\prod _{j<k} r_{jk}^{|\alpha|}$ is nothing but a
pseudo gauge transformation factor, whose parameter is the real part of
the analytic
function $|\alpha|\sum_{j<k}\ln z_{jk}$. The   imaginary part of the same
analytic function
is precisely the singular gauge transformation parameter which defines
the anyonic $N$-body  vector potential $\vec A(\vec
r_i)={\alpha\over e} \vec {\partial}_i\sum_{j<k}\phi_{jk}$. It is not difficult
to
realize that,
due to the
Cauchy-Riemann relations, $\sum_i\vec {A}^2(\vec r_i)$  indeed
disappears  in $\tilde{H}$ \cite{Ouvry92} \cite{dd}.

To gain a more complete understanding of the
singular perturbative algorithm,
let us now try to see how it applies in a regular case. We
consider first
a smeared flux tube version of the singular Aharonov-Bohm problem -possible
generalizations involve flux tubes of finite
size-
and we
concentrate on a flux smeared over a certain
region of space, with a given profile.
The effective change of statistics of the
particles then depends on the distance between
them \cite{Mash93}. Thus consider
the vector potential
\be \vec A(\vec r)={\alpha\over e}{\vec k \times \vec r\over r^2}
\varepsilon(r)\ee
where $\varepsilon(r)$ satisfies the boundary
conditions
$\varepsilon(\infty) = 1$ (hence at large distances
one has effectively anyons with statistics
$\alpha$) and $\varepsilon(0)=0$,
in order to avoid singularities at the origin. The physical
meaning of $\varepsilon(r)$ is rather obvious :
$\Phi(r) = {2\pi{\alpha\over e}\varepsilon(r)}$
is the flux through a circle of radius $r$,
and
\begin{equation}
B(r)=\frac{\alpha}{er} \frac{d\varepsilon(r)}{dr}
\label{magnf}
\end{equation}
is the magnetic field profile of the smeared flux tube.

The Hamiltonian now reads
\begin{equation}
{\cal H} = {1\over 2m}(-\frac{\partial^2}{\partial r^2}
- \frac{1}{r}\frac{\partial}{\partial r}
- \frac{1}{r^2}\frac{\partial^2}{\partial\phi^2}
+ \frac{2i\alpha\varepsilon(r)}{r^2}\frac{\partial}{\partial\phi}
+ \frac{\alpha^2\varepsilon^2(r)}{r^2})+ {1\over 2}m\omega^2r^2
\label{Hamdds}
\end{equation}
In the problem at hand, there is always a characteristic
parameter $R$, which is essentially the size of the
flux tube, such that $\varepsilon(r) \sim 1$ for $r \gg R$.
All the results of the ideal anyon model should be recovered
in the limit $R \rightarrow 0$.
Since the Hamiltonian ${\cal H}$ tends to
$H$ in this limit, the problem of perturbation
theory does manifest itself for ${\cal{H}}$.
Indeed, the first order correction to the ground state energy
\begin{equation}
\left\langle \psi^{(0)}_{00} \right|\frac{\alpha^2\varepsilon^2(r)}{2mr^2}
\left| \psi^{(0)}_{00}\right\rangle
= \int\limits_0^{\infty}\frac{\omega\alpha^2\varepsilon^2(r)}{r}  e^{-m\omega
r^2} dr
\label{corrdds}
\end{equation}
is finite, but diverges
as $R \rightarrow 0$, whereas it should
tend to $|\alpha|\omega$.

What stands,  in this smeared case,
in place of the singular A-B perturbative algorithm?
Let us recall that for ideal anyons, the magnetic
field inside the singular flux tube is
$B(r)={2\pi{\alpha\over e}}\delta^2({\vec r})$.
The ${\pi|\alpha|\over m}\delta^2(\vec r)$ contact term
added to $H$ may be interpreted as the coupling to the singular magnetic
field of a
magnetic moment $\mu$ associated to the
particle\footnote{ Such magnetic moment couplings have
already been introduced in the anyon model \cite{stern}, as relics of a
relativistic formulation, but were shown to be associated to attractive
$\delta^2$ interactions.}
\begin{equation}
\mu = -{e\over 2m}{\alpha\over|\alpha|}
\label{magnmom}
\end{equation}
Coming back to the smeared flux case, this suggests to introduce
\begin{equation}
{\cal H}' = {\cal H} -{ \mu } B(r)
\label{Hddspr}\ee
corresponding to the magnetic moment coupling
\be -{e\over 2m}g {\sigma_3\over 2}B(r)\ee
with the gyromagnetic factor $g=2$.
What is now the appropriate generalization of
the ansatz (\ref{ansatz}) for the wave function? In the singular case,
the idea  was to extract the short distance ground state
behavior. It happens that the ground state wave functions
for a 2-dimensional particle with the gyromagnetic factor $g=2$ in
a magnetic field $B$ are (up to a holomorphic function) \cite{AC}
\begin{equation}
\psi_{00} = e^{-2m \mu a(r) }
\label{grst}
\end{equation}
where $a(r)$ is such that
\begin{equation}
\Delta a(r) = B(r)
\label{lapl}
\end{equation}
In \cite{AC}, spin-$\frac{1}{2}$
particles have been considered, altogether with a
 Pauli Hamiltonian viewed as the nonrelativistic
limit of the relativistic Dirac
Hamiltonian. In the present context, however, spin is an additional
degree of freedom
simply introduced by hand.
Taking into account (\ref{magnf}), one has
\be a(r) = {\alpha\over e}\int_{0}^{r} \frac{\varepsilon(r')}{r'} \, dr'\ee
and the generalized ansatz is
\begin{equation}
\psi(r,\phi) = \exp \left[ \int\limits_0^r
|\alpha|\frac{\varepsilon(r')}{r'} \, dr' \right] \tilde{\psi}(r,\phi).
\label{ansatzdds}
\end{equation}
Transforming $\cal{H}'$, one obtains
\begin{equation}
\tilde{\cal H} ={1\over 2m}(
-\frac{\partial^2}{\partial r^2}
-\frac{1}{r}\frac{\partial}{\partial r}
-\frac{1}{r^2}\frac{\partial^2}{\partial\phi^2}
+ \frac{2i\alpha\varepsilon(r)}{r^2}\frac{\partial}{\partial\phi}
- \frac{2|\alpha|\varepsilon(r)}{r}\frac{\partial}{\partial r}) +{1\over
2}m\omega^2r^2
\label{Htildedds}
\end{equation}
where the ${\vec {A}}^2$ have  again disappeared.

In a sense, coming back to the singular case,
one has now at hand a clearer point of view
on certain subtleties associated with the
contact term,  and also a more precise understanding
of the ansatz for the redefinition of the wave function.  A ``naive''
$R \rightarrow 0$ limit $\varepsilon(r) \equiv 1$ would imply that {\it both\/}
${\cal H}$ and ${\cal H}'$ would coincide
with $H$. However, if one insists on the
non-singular boundary condition $\varepsilon(0)=0$, then
in the limit $R\to 0$ one should rather take $\varepsilon(r) = \eta(r)$,
where $\eta(r)$ is the step function.  Then ${d\varepsilon(r)\over dr}
=\delta(r)$, and ${\cal H}'$ coincides with $H'$, not $H$.
Here, to ignore the difference between unity and the step function
would be the same as, say, to consider that $\Delta \ln r = 0$,
 rather than
$\Delta \ln r = 2\pi\delta^2({\vec r})$, thus
``losing'' the $\delta^2(\vec r)$ contact term \cite{Ouvry94}. Note also
that once the correct ansatz is made, i.e. once one works with
${\tilde {\cal H}}$, this subtlety does not anymore play any role : in the
limit
$R\to 0$,
it does not matter whether to put $\varepsilon=1$ or
$\varepsilon=\eta(r)$ to get (\ref{Htilde}),
since the correct short-distance behavior has
already been properly taken into account.

Generalizing further, consider now the Hamiltonian
\be 2m {\cal H}'_{\pm}=(\vec p -e\vec A)^2 \mp e(\partial_1 A_2-\partial_2
A_1)\ee
and go to the Coulomb gauge
$A_1=-\partial_2 a, \quad A_2=\partial_1 a$. In 2 dimensions,  this is
a general choice of gauge.
One has
\be 2m {\cal H}'_{\pm}=-\Delta -2ei\epsilon_{ij}\partial_ja\partial_i+e^2(\vec
\partial a)^2\mp e\Delta a\ee
Redefine\footnote{Note that  the inverse transformation
 $\psi=e^{\pm ea}{\chi}$ leads to the Fokker-Planck equation
associated to ${\cal H}'_{\pm}$
$$ -\Delta \chi+\partial_i(\chi K_i)=E\chi $$
$$ K_i=\pm 2e(\partial_ia+i\epsilon_{ij}\partial_j a)$$}
$\psi=e^{\mp ea}\tilde{\psi}$.
If $\tilde{\psi}=\tilde{\psi}(z)$ (respectively
$\tilde{\psi}=\tilde{\psi}(\bar z)$), then $\psi$ is
the zero energy ground state wave function of $H_+$
(respectively $H_-$). Otherwise, one gets, acting on $\tilde{\psi}$,
\be 2m \tilde{{\cal H}}_{\pm}=-\Delta -2ei\epsilon_{ij}\partial_ja\partial_i
\pm 2e\partial_ia\partial_i\ee

The connection with what has been said above is transparent if one
specializes to the rotationally invariant case
$ \epsilon_{ij}\partial_ja\partial_i
=-{da(r)\over dr}\partial_{\phi}$. Focusing on
the s-wave sector, only the term $\pm 2e\partial_ia\partial_i$ contributes
to the energy shift
\be E-E_0= \pm{e\over
m}\int\psi^{(0)}_{00}\partial_ia\partial_i\psi^{(0)}_{00}d^2\vec r=
\pm {e\over 2m}\int\partial_i|\psi^{(0)}_{00}|^2\partial_i ad^2\vec r=\pm
{e\over 2m}\int|\psi^{(0)}_{00}|^2\Delta
a d^2\vec r \ee

If one wishes to generalize to the $N$-body case, one starts from
\be 2m {\cal H}'_{\pm}=\sum_{i=1}^N(\vec p_i-e\vec A_i)^2\mp e B(\vec r_i)\ee
with
\be \vec A(\vec r_i)=-\vec \partial_i \times \sum_{j<k} a(r_{jk}),\quad B(\vec
r_i)=
\Delta_i \sum_{j<k} a(r_{jk})\ee
and redefines
\be \psi=\prod_{j<k}e^{\pm ea(r_{jk})}\tilde{\psi}\ee
to get a Hamiltonian without 3-body
interactions, exactly as in the $N$-anyon case.

To conclude, and as an explicit illustration, let us carry out the
calculation in the simple case where the
magnetic field is uniform within a circle of
radius $R$. One has
\begin{equation}
\varepsilon(r) = \left\{ \begin{array}{lc}
\frac{r^2}{R^2} \: , & r \leq R, \\ \\
1 \: , & r \geq R. \end{array} \right.
\label{model}
\end{equation}
The first-order correction from the last term
of ${\cal H}'$ in (19) is
\be
\left\langle \psi^{(0)}_{00} \right| \frac{|\alpha|}{2mr}
\frac{d\varepsilon(r)}{dr} \left| \psi^{(0)}_{00} \right\rangle
=\frac{1-\exp(-q)}{q} |\alpha|\omega,
\ee
where
\begin{equation}
q={m\omega} R^2
\label{q}
\end{equation}
is the squared ratio of the flux tube radius to the
length scale of the harmonic potential: The particle is
well outside the flux tube if $q \ll 1$. In the limit
$q \rightarrow 0$, the exact result (\ref{E0})
is recovered. Alternatively, one may proceed
with the Hamiltonian (\ref{Htildedds}) to get
\be
\left\langle \psi^{(0)}_{00} \right|
- \frac{|\alpha|\varepsilon(r)}{mr}\frac{\partial}{\partial r}
\left| \psi^{(0)}_{00} \right\rangle
=\frac{1-\exp(-q)}{q} |\alpha|\omega,
\ee
the same answer as above.

In conclusion, hard core boundary prescriptions in the singular
A-B/anyon cases can be naturally
understood in the context of Aharonov-Casher Hamiltonians for spin 1/2
particles coupled to 2-d magnetic field, with the gyromagnetic factor
$g=2$.

Acknowledgements :
S.M. thanks the theory division of the IPN
for the hospitality extended to him during the initial
stage of this work.

\end{document}